# Synthetic Weyl points in generalized parameter space and their topological properties

Qiang Wang,[1*] Meng Xiao,[2,3*] Hui Liu,[1] Xiangang Wan,[1]
Shining Zhu,[1] and C. T. Chan[2]

[1]*National Laboratory of Solid State Microstructures, School of Physics, Collaborative Innovation Center of Advanced Microstructures, Nanjing University, Nanjing 210093, China*
[2]*Department of Physics, Institute for Advanced Study, the Hong Kong University of Science and Technology, Clear Water Bay, Kowloon, HongKong*
[3]*Department of Electrical Engineering, and Ginzton Laboratory, Stanford University, Stanford, California 94305, USA*

* These authors contribute equally. Correspondence and requests for materials should be addressed to C.T.C. (email: phchan@ust.hk ) or to H.L. (email: liuhui@nju.edu.cn).

Weyl fermions[1] do not appear in nature as elementary particles, but they are now found to exist as nodal points in the band structure of electronic[2,3,4,5,6] and classical wave crystals[7,8,9,10,11,12,13,14]. Novel phenomena such as Fermi arcs[3,6,15] and chiral anomaly[16] have fueled the interest of these topological points which are frequently perceived as monopoles in momentum space. Here, we demonstrate that generalized Weyl points can exist in a parameter space and we report the first observation of such nodal points in one-dimensional photonic crystals in the optical range. The reflection phase inside the band gap of a truncated photonic crystal exhibits vortexes in the parameter space where the Weyl points are defined and they share the same topological charges as the Weyl points. These vortexes guarantee the existence of interface states, the trajectory of which can be understood as "Fermi arcs" emerging from the Weyl nodes.

Great efforts have been devoted to investigate the intriguing phenomena associated with Weyl points[1, 3, 4, 6, 16], such as the Fermi arc surface states[15, 17] and chiral anomaly[16] associated with electronic systems. Besides electronic systems, Weyl points have also been found in photonic [7, 10, 12, 13, 18, 19], acoustic[11] and plasmon[14] systems. To-date, Weyl points are frequently identified as "momentum space magnetic monopoles", meaning that they are sources or sinks of Berry curvature defined in the momentum space. As such, Weyl points are perceived as topological nodal points in 3D momentum space defined by Block momentum coordinates $k_x, k_y$ and $k_z$. Here, we show that topological nodal points can also be meaningfully defined in a generalized phase space in which some of the dimensions are not Bloch momentum, and yet their topological characters can give analogues of Fermi-arc-like surface states, in much the same way as ordinary Weyl points defined in 3D k-space. We will use photonic crystals to realize the ideas of generalized Weyl points experimentally in the optical frequency regime. We note that realizing Weyl point phenomena in optical frequency is by itself as a great challenge, as Weyl crystals tend to be structurally complex and hence difficult to fabricate in optical frequencies.

The Weyl Hamiltonian can be written as $H = \sum_{i,j} k_i v_{i,j} \sigma_j$ [20], where $v_{i,j}$, $k_i$ and $\sigma_j$ with $i = x, y, z$ representing the group velocities, wave vectors and the Pauli matrices, respectively. Compared with two-dimensional (2D) Dirac points, Weyl points are stable against perturbations that keep the wave vectors as good quantum numbers[7]. Each Weyl point has its associated topological charge given by the Chern number of a closed surface enclosing it[5]. Though Weyl points are usually defined in 3D momentum space, Weyl physics have been discussed in synthetic dimensions very recently[21]. Instead of using all three components of the wave vector, synthetic dimensions can be used to replace one or two of the momentum components. Synthetic dimensions have attracted attention recently due to their ability to realize physics in higher dimensions[22, 23, 24, 25] and to simplify experimental design[21]. As we will show below, such generalizations enable the investigation of Weyl points in optical regime while preserving the standard Weyl point characters such as the associated topological charges[12, 13] and robustness against variation of the parameters[7]. In this work, we replace two wave vector components with two independent geometric parameters (which form a parameter space) in the Weyl Hamiltonian, and our design can be realized with simple 1D photonic crystals (PCs).

We shall see that the reflection phase of a truncated PC exhibits vortex structure[26] in the parameter space around a synthetic Weyl point. These vortexes carry the same topological charges as the corresponding Weyl points. The reflection phase along any loop in parameter space encircling the center of the vortex (also the position of the Weyl point) varies continuously from $-\pi$ to $\pi$. This property guarantees the existence of interface states in the boundary separating the PC and a gapped material such as a reflecting substrate [27, 28] independent of the properties of the substrate. These interface states connect Weyl points with opposite charges or extend from one Weyl point to the boundary of the parameter space, which can be regarded as analogues of chiral edge states in Weyl semimetal[4, 6, 17]. Meanwhile, we also observe the topological transition[7] by varying an additional parameter of the PCs.

To illustrate the idea of synthetic Weyl points in generalized parameter space, we consider a one-dimensional (1D) PC consisting of 4 layers per unit cell, as shown in the inset in Fig. 1a. In our experiments, the first and third layers (blue) are made of H$_f$O$_2$ with refractive index $n_a$ =2.00, and the second and forth layers (red) are S$_i$O$_2$ with refractive index $n_b$ =1.45. The thickness of each layer is given by:

$$\begin{aligned} d_{a1} &= (1+p) \cdot d_a \\ d_{b1} &= (1+q) \cdot d_b \\ d_{a2} &= (1-p) \cdot d_a \\ d_{b2} &= (1-q) \cdot d_b \end{aligned} \quad (1)$$

where we define two parameters, $p$ and $q$, which both belong to [-1,1]. In the experiments, we choose $d_a = 0.323 um$, $d_b = 0.240 um$. The total optical length $L$ inside the unit cell is a constant $2(n_a d_a + n_b d_b)$ for the whole $p$-$q$ space as sketched in Fig. 1a. The structural parameters $p$ and $q$, together with one Bloch wave vector $k$ construct a 3D parameter space in which we study the Weyl physics.

Let us start with the PCs with only two layers inside each unit cell, say, H$_f$O$_2$ with thickness $d_a$, and S$_i$O$_2$ with thickness $d_b$. The band dispersion is plotted in Fig. 1b in red. A four-layer PC with parameters $p=0$ and $q=0$ just doubles the length of each unit cell and folds the Brillouin zone. The dispersion of this four-layer PC is shown in Fig. 1b in blue. Such artificial band folding gives us a linear cross along the wavevector direction. Away from the point $p=0$ and $q=0$, the degeneracy introduced by the band folding is lifted and a band gap emerges. Fig. 1c shows the band dispersions in the $p$-$q$ space with $k=0.5k_0$, here $k_0 = \pi/(d_a+d_b)$. Two bands form a conical intersection indicating that band dispersion is linear along all the directions. To characterize this point, we derive an effective Hamiltonian

for parameters around this degenerate point (see Supplementary Material Sec. I):

$$H = c_1 p\sigma_x + c_2 p\sigma_y + c_3 q\sigma_x + c_4 q\sigma_y + c_5 \xi_k \sigma_z, \qquad (2)$$

where $\xi_k = (k-0.5k_0)/k_0$, $c_1 = 0.057$, $c_2 = 0.091$, $c_3 = 0.079$, $c_4 = -0.050$ and $c_5 = 1.956$. This Hamiltonian possesses a standard Weyl Hamiltonian form and is located at $(p,q,k) = (0,0,0.5k_0)$ with "charge" -1 according to the usual definition. The topological charge of this Weyl point can also be numerically verified. Here we adapt the method in ref 5 to calculate the Weyl point charge. As shown in the inset in Fig. 1d, the Berry phases are defined on a spherical surface with fixed azimuthal angle $\theta$ (red circle). We then track the evolution of the Berry phases as functions of $\theta$, as shown in Fig. 1d. According to Fig. 1d, the Chern number for the lower band is -1, while +1 for the upper band, indicating the charge of the Weyl point here is -1, which is consistent with the effective Hamiltonian. (Please refer to Supplementary Material Sec. II for more details).

In addition to the Weyl point constructed above, we also find Weyl points on higher bands and at different positions in the parameter space. In Fig. 2, we show all the Weyl points on the lower five bands at either $k = 0.5k_0$ or $k = 0$, and the inserts mark the locations and charges of these Weyl points. The Weyl point between band 1, the lowest band, and band 2 (Fig. 2a) with charge -1 has already been discussed above. There are two Weyl points with charge -1 between band 2 and band 3. Weyl points with charge +1 appear on higher bands. Our system possesses the symmetry $\varepsilon(p,q,x) = \varepsilon(-p,q,-x) = \varepsilon(p,-q,-x)$, where $\varepsilon$ is the permittivity and $x$ represents the real space position. This symmetry requires that once there is a Weyl point at $(p_0, q_0)$ at a particular frequency, there are other Weyl points at $(-p_0, q_0)$, $(p_0, -q_0)$ and $(-p_0, -q_0)$, and these Weyl points all possess the same topological charge. We note that while the total charges of Weyl points must vanish in periodic systems[29], such constraint does not apply here as the parameter space is not closed[12].

We then consider the reflection phase of a normal incident wave when the PC is semi-infinite. When the working frequency of the incident wave is chosen to be the frequency of the Weyl point (between band 1 and band 2), except for the $p = q = 0$ point, the working frequency is inside the bandgap for all other $p$ and $q$ values. Hence the reflection coefficient can be written as $r = e^{i\psi}$, with $\psi$ being a function of $p$ and $q$. In Fig. 3a, we show the reflection phase in the whole $p$-$q$ space, where the truncation boundary is at the center of the first layer. The reflection phase distribution shows a vortex structure, with the Weyl point located at the vortex center. Gray arrows in Fig. 3a denote the direction of the phase gradient. The topological charge of this vortex is given by the winding number of the phase gradient, which is the same as that of the Weyl point. To elaborate more on this point, we choose two loops: one encircles the Weyl point (white circle in Fig. 3a), and the other do not (black circle in Fig. 3a). The reflection phases along these two loops are plotted in Fig. 3b with the blue and red curves, respectively. The reflection phase along the white loop in Fig. 3a decreases with the polar angle $\varphi$, and picks up a total change of $-2\pi$ after each circling. Similarly, if the charge of the Weyl point is +1, then the total phase change is $2\pi$ for each circling. A proof with the effective Hamiltonian can be found in the Supplementary Material Sec. III. The reflection phase along the black circle in Fig. 3a picks up a zero total change along each loop. We emphasis that the vortex structure and the charge of the vortex are independent of the position where we truncate the PCs. (See the Supplementary Material Sec. III)

The vortex structure guarantees the existence of interface states between the PCs with Weyl points and reflecting substrates independent of the properties of the substrate. The existence of surface states is given by

$$\psi_{PC} + \psi_S = 2m\pi, \quad m \in \mathbb{Z}, \tag{3}$$

where $\psi_{PC}$ and $\psi_S$ represent the reflection phases of the PC and the reflecting substrate, respectively. As the reflection phase on the loop circling the Weyl point covers the whole region $[-\pi, \pi)$, no matter what the reflection phase of the reflecting substrate is, Eq. (3) can always be satisfied at one polar angle. If we consider loops with different radii, then the surface states form a continuous trajectory that begins from the Weyl point. The trajectory of the surface states ends either at another Weyl point with opposite charge or the boundary of the parameter space. The behavior of surface states connecting Weyl points with opposite charges is similar to the Fermi arc [2, 3, 16, 17] in Weyl semimetals. The most prominent character of Weyl semimetals is perhaps the existence of the Fermi arc surface states connect two Weyl points with opposite charges. There is however a difference: the Fermi arc starts and ends with Weyl points in periodic system, while the surface states in our system can connect Weyl points with the boundary of the parameter space due to the non-vanishing of total charge of the Weyl points inside the parameter space. As an example, here we consider the Weyl points on the fourth and fifth bands as shown in Fig. 2d. There are in total 8 Weyl points: 6 Weyl points with charge -1 and the remaining 2 with charge +1. As our system possesses no external symmetry which can relate two Weyl points with opposite charges, the frequency of the Weyl point with positive charge is higher than that of the Weyl points with negative charge. We choose the working frequency to be at the frequency of the Weyl point with a positive charge (303THz) and the PCs are truncated at the center of the first layer. In Fig. 4a, we show the reflection phase in the parameter space, where gray areas within black dashed lines mark the regions of the bulk band. If the truncated PCs are coated with silver films, then there exist interface states which satisfy Eq. (3). The reflection phase of silver is measured to be $-0.95\pi$ at 303THz. White dashed lines in Fig. 4a shows the trajectories of surface states. Indeed besides surface states connecting Weyl points with different charges, there are also trajectories of surface states terminated at the boundary of our parameter space. We also perform experiments to verify the results, we choose four points (positions marked with cyan triangles in Fig. 4a) to experimentally demonstrate the existence of the surface states, with samples consist of ten periods, and the results are shown in Fig. 4b with red crosses, which matches well with the numerical data (cyan dashed line). We emphasize here that the existence of surface states is "robust" to the property of the reflecting substrates: we can always find trajectories of interface states that link the two Weyl points with opposite charges no matter what the reflecting substrate is.

We can also define other parameters to extend to higher dimensions, and such extensions enable the studies of phenomena which exist only in higher dimension spaces[24]. As an example, we can define another parameter "$R$" as ratio $R = n_a d_a / (n_a d_a + n_b d_b)$, which belongs (0, 1). With the variation of this parameter, we now observe topological transitions of the band structures. In Fig. 5, we show the dispersion of the fourth and fifth bands gap with $k = 0$ for different values of $R$. Insets show the positions of the Weyl points (black circles) and degenerate lines (dashed lines). When the $R$ = 0.25 (Fig. 5c), 0.5 (Fig. 5e) and 0.75 (Fig. 5g), there exist only the degenerate lines, which act as analogues of the nodal lines in semimetal[30]. These nodal lines act as topological transition points between configurations with different Weyl points. Though the net charge does not vanish, the total charge in the parameter space keeps constant.

In summary, we show the existence of Weyl points in the parameter space. In the specific example of dielectric superlattices, the reflection phase of the semi-infinite multilayered PC shows a vortex structure with the same topological charge as the synthetic Weyl points defined in the parameter space, and such character guarantees the existence of interface states. We emphasize that in general, interface states may or may not exist in the boundary between a 1D PC and a reflecting substrate, and the Weyl points here provide us a deterministic scheme to construct boundary modes between multi-layered PCs and reflecting substrates of arbitrary reflection phases[28]. We further note that the topological notions works in high frequency gaps where the PCs cannot be described by the effective medium theory. This work will also pave a new direction to explore the physics in topological theory in higher dimensions[24]. In addition, the reflection phase vortex gives a flexible way to manipulate the electromagnetic wave, such as generating vortex beam and controlling the reflection direction. With more parameters involved, we can also construct Weyl points with higher charge[13, 19, 31].

# Method Section

**Sample fabrication.** The PCs are fabricated by Electron Beam Evaporation on the substrate made of K9 glass. Before evaporation, we use an acid solution to clean the substrate. During the evaporation, the pressure in the chamber is kept below $2\times10^{-3}$ Pa, and the temperature is maintained at 90℃. The silver film is fabricated by ion beam sputtering, the thickness is controlled at about 20nm after sputtering for 2.5min. The thickness uncertainty of each layer in the fabrication is below 10nm for all the PCs., The number of unit cells are 10, and $d_a = 0.323um$ and $d_b = 0.240um$ are fixed for the 4 PCs in Fig. 4b. The thickness of layers for these 4 PCs are given by $(p,q) = (0.50, 0.26)$, $(0.56, 0.30)$, $(0.70, 0.38)$ and $(0.78, 0.36)$ respectively.

# Figures

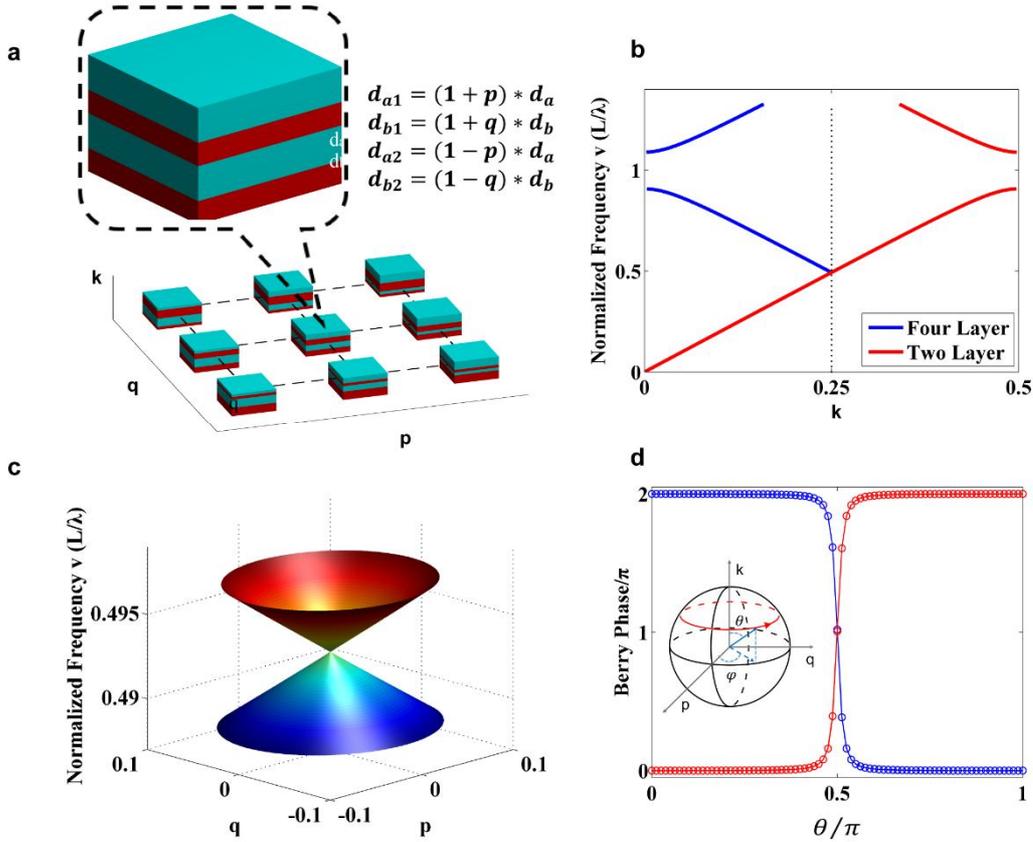

**Figure 1 | Realizations of Weyl points in the parameter space. a**, Photonic crystals (PCs) with different $p, q$ values. The $p, q$ form a parameter space. The inset shows an example of one unit cell, where the first and the third layers are made of $H_fO_2$ (blue), and the second and the forth layers are made of $S_iO_2$ (red), the thickness of each layer depends on the position in the $p$-$q$ parameter space. **b**, The band dispersion of PCs with different unit cells, red color for a PC with only two layers, and blue color for a PC which just doubles the unit cell used for the red line. Here $d_a = 0.323 um$ and $d_b = 0.240 um$. **c**, The dispersion of PCs in the $p$-$q$ space with $k=0.5\,k_0$. Here two bands form a conical intersection. **b** and **c** together show that the band dispersions are linear along all the directions around the degeneracy point, indicating that it is an analogue of Weyl point. **d**, Berry phases defined on the spherical surface with fixed $\theta$ (see the inset), where blue

and red represent Berry phases on the lower and the upper band, respectively. The radius of the sphere is 0.001.

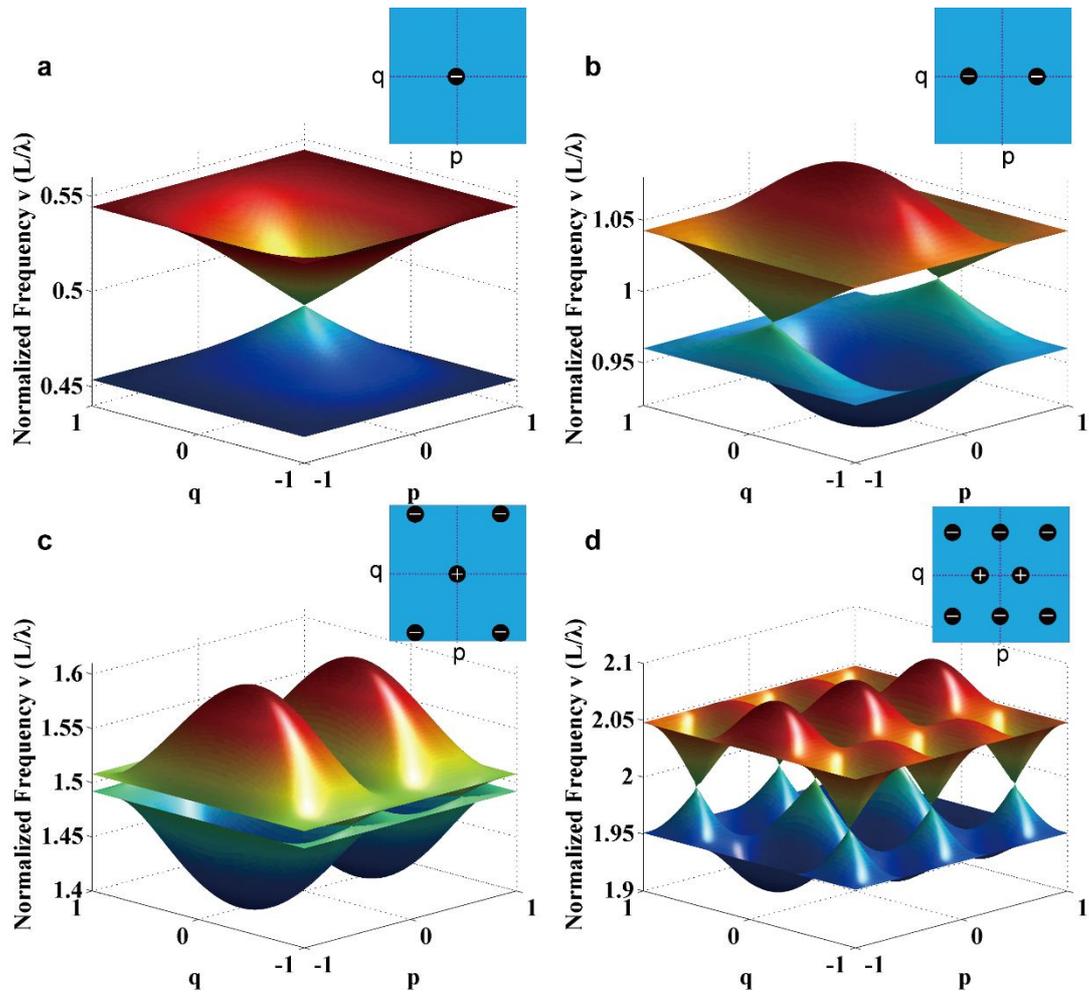

**Figure 2 | The band dispersions in the parameter space at different k points. a** and **c,** the band dispersions of the lowest four bands at $k=0.5k_0$. **b** and **d,** the band dispersions of band 2 to band 5 at $k=0$. The conical intersections in **a-d** all correspond to Weyl points with positions and charges ("+" for charge +1, "-" for charge -1) marked in the insets, in which the dashed lines denote $p=0$ or $q=0$. Here we use the same parameters as those in Fig. 1.

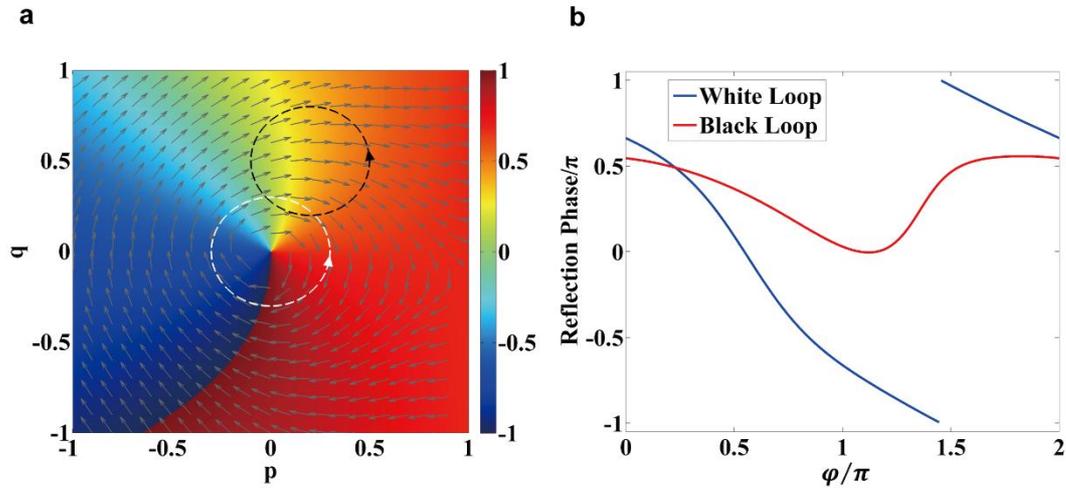

**Figure 3 | The reflection phase in the $p$-$q$ space. a**, The reflection phase in the $p$-$q$ space at the frequency of the Weyl point in Fig. 1. The gray arrows denote the directions of the reflection phase gradient. The reflection phase shows a vortex structure with charge -1. The white and black circles mark two paths along which the reflection phases are shown in **b** with the blue and red curves, respectively. The radii of these two circles are 0.3. The white circle centers at $(p, q) = (0, 0)$, while the black circle centers at $(p, q) = (0.2, 0.5)$. Here we use the same parameters as those in Fig. 1.

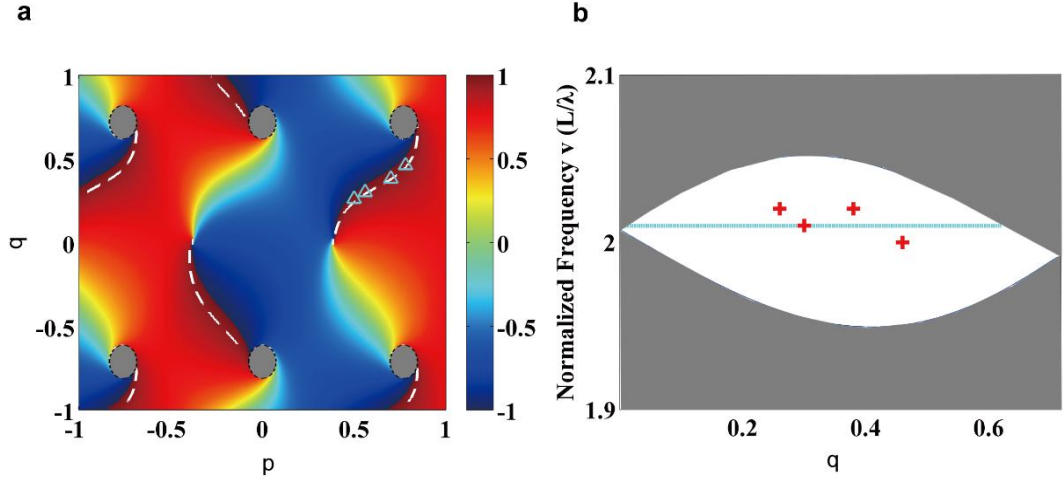

**Figure 4 | The 'Fermi arc' in the parameter space. a**, The reflection phase of the PCs at the frequency of Weyl points with charge +1 with its dispersion shown in Fig. 2d. The gray regions encircled by the dashed black lines represent the bulk band regions at the working frequency. The white dashed lines show the trajectories of the interface states of a system consisting of a semi-infinite PC coated by a silver film, where the semi-infinite PC is truncated at the center of the first layer. These surface state trajectories are analogues of Fermi arc states. The triangles mark the $p$, $q$ values of the four samples in the experiment. **b**, The cyan line represents the working frequency used in **a**, and the red crosses label the experimental results, the gray region mark the projection of the bulk band as a function of $q$. The number of unit cells are 10, and $d_a = 0.323 um$ and $d_b = 0.240 um$ are fixed for the 4 PCs in **b**. The thickness of layers for these 4 PCs are given by $(p, q) = (0.50, 0.26)$, $(0.56, 0.30)$, $(0.70, 0.38)$ and $(0.78, 0.36)$ respectively.

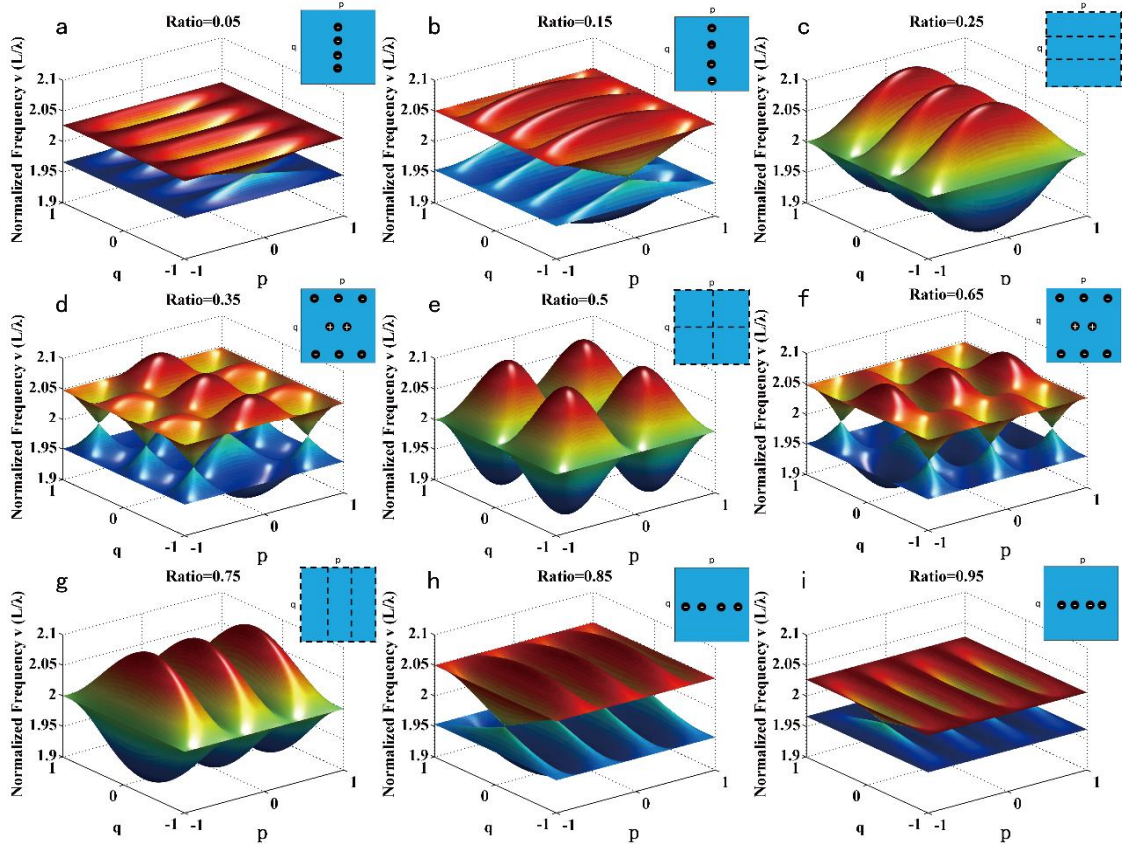

**Figure 5 | Topological transitions occur as we change the parameter Ratio defined in text. a-i,** The band structures of band 4 and band 5 with different Ratios at $k$=0. Insets show the positions of the Weyl points ("+" for positive charge and "-" for negative charge) and "nodal lines" (black dashed lines).

**Reference**

1. Weyl H. Elektron und Gravitation. I. *Zeitschrift für Physik* 1929, **56**(5)**:** 330-352.

2. Wan X, Turner AM, Vishwanath A, Savrasov SY. Topological semimetal and Fermi-arc surface states in the electronic structure of pyrochlore iridates. *Physical Review B* 2011, **83**(20)**:** 205101.

3. Xu S-Y, Belopolski I, Alidoust N, Neupane M, Bian G, Zhang C*, et al.* Discovery of a Weyl fermion semimetal and topological Fermi arcs. *Science* 2015, **349**(6248)**:** 613.

4. Lv BQ, Weng HM, Fu BB, Wang XP, Miao H, Ma J*, et al.* Experimental Discovery of Weyl Semimetal TaAs. *Physical Review X* 2015, **5**(3)**:** 031013.

5. Soluyanov AA, Gresch D, Wang Z, Wu Q, Troyer M, Dai X*, et al.* Type-II Weyl semimetals. *Nature* 2015, **527**(7579)**:** 495-498.

6. Deng K, Wan G, Deng P, Zhang K, Ding S, Wang E*, et al.* Experimental observation of topological Fermi arcs in type-II Weyl semimetal MoTe2. *Nat Phys* 2016, **advance online publication**.

7. Lu L, Fu L, Joannopoulos JD, Soljacic M. Weyl points and line nodes in gyroid photonic crystals. *Nat Photon* 2013, **7**(4)**:** 294-299.

8. Lu L, Wang Z, Ye D, Ran L, Fu L, Joannopoulos JD*, et al.* Experimental observation of Weyl points. *Science* 2015, **349**(6248)**:** 622.

9. Chen W-J, Xiao M, Chan CT. Experimental observation of robust surface states on photonic crystals possessing single and double Weyl points.   *ArXiv e-prints*; 2015.

10. Jorge B-A, Ling L, Liang F, Hrvoje B, Marin S. Weyl points in photonic-crystal superlattices. *2D Materials* 2015, **2**(3)**:** 034013.

11. Xiao M, Chen W-J, He W-Y, Chan CT. Synthetic gauge flux and Weyl points in acoustic systems. *Nat Phys* 2015, **11**(11)**:** 920-924.

12. Xiao M, Lin Q, Fan S. Hyperbolic Weyl Point in Reciprocal Chiral Metamaterials. *Physical Review Letters* 2016, **117**(5)**:** 057401.

13. Chang M-L, Xiao M, Chen W-J, Chan CT. Multi Weyl Points and the Sign Change of Their Topological Charges in Woodpile Photonic Crystals.   *ArXiv e-prints*; 2016.

14. Gao W, Yang B, Lawrence M, Fang F, Beri B, Zhang S. Photonic Weyl degeneracies in magnetized plasma. *Nat Commun* 2016, **7**.

15. Xu S-Y, Alidoust N, Belopolski I, Yuan Z, Bian G, Chang T-R, *et al.* Discovery of a Weyl fermion state with Fermi arcs in niobium arsenide. *Nat Phys* 2015, **11**(9): 748-754.

16. Hosur P, Qi X. Recent developments in transport phenomena in Weyl semimetals. *Comptes Rendus Physique* 2013, **14**(9–10): 857-870.

17. Chang T-R, Xu S-Y, Chang G, Lee C-C, Huang S-M, Wang B, *et al.* Prediction of an arc-tunable Weyl Fermion metallic state in MoxW1-xTe2. *Nat Commun* 2016, **7**.

18. Wang L, Jian S-K, Yao H. Topological photonic crystal with equifrequency Weyl points. *Physical Review A* 2016, **93**(6): 061801.

19. Chen W-J, Xiao M, Chan CT. Photonic crystals possessing multiple Weyl points and the experimental observation of robust surface states. *Nature Communications* 2016, **7**: 13038.

20. Fang Z, Nagaosa N, Takahashi KS, Asamitsu A, Mathieu R, Ogasawara T, *et al.* The Anomalous Hall Effect and Magnetic Monopoles in Momentum Space. *Science* 2003, **302**(5642): 92.

21. Lin Q, Xiao M, Yuan L, Fan S. Photonic Weyl Point in a Two-Dimensional Resonator Lattice with a Synthetic Frequency Dimension. *Nat Commun* 2016.

22. Yuan L, Shi Y, Fan S. Photonic gauge potential in a system with a synthetic frequency dimension. *Opt Lett* 2016, **41**(4): 741-744.

23. Price HM, Zilberberg O, Ozawa T, Carusotto I, Goldman N. Four-Dimensional Quantum Hall Effect with Ultracold Atoms. *Physical Review Letters* 2015, **115**(19): 195303.

24. Lian B, Zhang S-C. Five-dimensional generalization of the topological Weyl semimetal. *Physical Review B* 2016, **94**(4): 041105.

25. Boada O, Celi A, Latorre JI, Lewenstein M. Quantum Simulation of an Extra Dimension. *Physical Review Letters* 2012, **108**(13): 133001.

26. Yu N, Genevet P, Kats MA, Aieta F, Tetienne J-P, Capasso F, *et al.* Light Propagation with Phase Discontinuities: Generalized Laws of Reflection and Refraction. *Science* 2011, **334**(6054): 333.

27. Xiao M, Zhang ZQ, Chan CT. Surface Impedance and Bulk Band Geometric Phases in One-Dimensional Systems. *Physical Review X* 2014, **4**(2): 021017.

28. Wang Q, Xiao M, Liu H, Zhu S, Chan CT. Measurement of the Zak phase of photonic bands through the interface states of a metasurface/photonic crystal. *Physical Review B* 2016, **93**(4): 041415.

29. Nielsen HB, Ninomiya M. Absence of neutrinos on a lattice. *Nuclear Physics B* 1981, **193**(1)**:** 173-194.

30. Fang C, Chen Y, Kee H-Y, Fu L. Topological nodal line semimetals with and without spin-orbital coupling. *Physical Review B* 2015, **92**(8)**:** 081201.

31. Fang C, Gilbert MJ, Dai X, Bernevig BA. Multi-Weyl Topological Semimetals Stabilized by Point Group Symmetry. *Physical Review Letters* 2012, **108**(26)**:** 266802.